\documentclass[twocolumn,pre,superscriptaddress]{revtex4}
\usepackage{graphics}
\usepackage{graphicx}
\usepackage{amsfonts}
\usepackage{textcomp}
\usepackage{amssymb}
\usepackage{mathrsfs}
\usepackage{amsmath}
\usepackage{color}
\usepackage{ulem}
\begin{document}

\title{Quantifying Microstructural Evolution via Time-Dependent Reduced-Dimension Metrics Based on Hierarchical $n$-Point Polytope Functions}

\author{Pei-En Chen\footnote{These authors contributed equally to this work.}}
\affiliation{Mechanical and Aerospace Engineering, Arizona State University, Tempe, AZ 85287}
\author{Rahul Raghavan\footnotemark[1]}
\affiliation{Materials Science and Engineering, Arizona State
University, Tempe, AZ 85287}
\author{Yu Zheng}
\affiliation{Department of Physics, Arizona State
University, Tempe, AZ 85287}
\author{Kumar Ankit}
\email[correspondence sent to: ]{kumar.ankit@asu.edu}
\affiliation{Materials Science and Engineering, Arizona State
University, Tempe, AZ 85287}
\author{Yang Jiao}
\email[correspondence sent to: ]{yang.jiao.2@asu.edu}
\affiliation{Materials Science and Engineering, Arizona State
University, Tempe, AZ 85287} \affiliation{Department of Physics,
Arizona State University, Tempe, AZ 85287}

\begin{abstract}
We devise reduced-dimension metrics for effectively measuring the distance between two points (i.e., microstructures) in the microstructure space and quantifying the pathway associated with microstructural evolution, based on a recently introduced set of hierarchical $n$-point polytope functions $P_n$. The $P_n$ functions provide the probability of finding particular $n$-point configurations associated with regular $n$-polytopes in the material system, and a special sub-set of the standard $n$-point correlation functions $S_n$ that effectively decomposes the structural features in the systems into regular polyhedral basis with different symmetry. The $n$-th order metric $\Omega_n$ is defined as the $\mathbb{L}_1$ norm associated with the $P_n$ functions of two distinct microstructures. By choosing a reference initial state (i.e., a microstructure associated with $t_0 = 0$), the $\Omega_n(t)$ set quantifies the evolution of distinct polyhedral symmetries and can in principle capture emerging polyhedral symmetries that are not apparent in the initial state. To demonstrate their utility, we apply the $\Omega_n$ metrics to a 2D binary system undergoing spinodal decomposition to extract the phase separation dynamics via the temporal scaling behavior of the corresponding $\Omega_n(t)$, which reveals mechanisms governing the evolution. Moreover, we employ $\Omega_n(t)$ to analyze pattern evolution during vapor-deposition of phase-separating alloy films with different surface contact angles, which exhibit rich evolution dynamics including both unstable and oscillating patterns. The $\Omega_n$ metrics have potential applications in establishing quantitative processing-structure-property relationships, as well as real-time processing control and optimization of complex heterogeneous material systems.

\end{abstract}
\maketitle

\section{Introduction}



The time-dependent behaviors of materials under extreme conditions (e.g., under ultra-fast cyclic thermal loading in additive manufacturing, or under chemically aggressive environment, or in the critical state of fracture) generally depend on coupled (non-equilibrium) processes that induce evolution of microstructural features on multi-length and time scales. Quantifying such microstructure evolution (i.e., 4D material behavior) is a crucial first step for understanding the physics governing the 4D material behaviors and the design and optimization of the material systems of interest.

One key challenge for microstructure quantification involves the hierarchy of structural disorder across multiple length scales \cite{torquato2002random}. Distinct from an crystalline or order system, which only requires a small number of ``representations'' (such as the set of lattice vectors) to uniquely and completely determine the structure, disordered systems are typically much more complex and require the specification of all degrees of freedom (e.g., the coordinates of all atoms in a metallic glass, or all pixel values for an image of a disordered composite material) for a complete description. Therefore, an alternative approach to quantification of disordered materials is to derive reduced dimension representations that statistically capture the key features of the systems, e.g., those crucial to determining the physical properties, instead of a precise description all of structural details \cite{jiao2007modeling, jiao2008modeling, jiao2009superior}.


Examples of quantitative representation of disordered systems include Gaussian random fields~\cite{roberts1997statistical}, geometric descriptors (e.g., grain/particle size and shape distribution)~\cite{wilding2011clustering, callahan2012quantitative, wang2012three, ratanaphan2014five}, spectral density functions~\cite{iyer2020designing, farooq2018spectral, chen2018designing}, and $n$-point correlation functions~\cite{niezgoda2008delineation, cecen2016versatile, choudhury2016quantification, okabe2005pore, fullwood2008microstructure, jiao2007modeling, jiao2008modeling, hajizadeh2011multiple, tahmasebi2013cross, tahmasebi2012multiple, xu2014descriptor, gerke2015improving, karsanina2018hierarchical, feng2018accelerating, gao2021efficient, gao2021ultra, malmir2018higher}, to name but a few.
The encoding process of these methods (e.g., extracting the representations from available structural or imaging data) are typically manually defined with clearly physical interpretations. However, due to the manual definitions, these representations often have limited degrees of freedom to approach completeness for arbitrary material systems \cite{jiao2010geometrical, gommes2012density, gommes2012microstructural}. On the other hand, machine learning (ML) techniques have recently been extensively applied in representation learning for complex disordered material systems. Most of these ML approaches propose either complete but non-explainable, or explainable but incomplete representations.
The former include purely data-driven generative models, e.g., restricted Boltzmann machines~\cite{cang2017microstructure}, variational autoencoders (VAE)~\cite{cang2018improving}, and generative adversarial networks (GAN)~\cite{yang2018microstructural,li2018transfer}, where a concise and near-complete representation is learned through microstructure samples, yet the encoders of which are composed of general-purpose neural networks and are non-explainable.

Among other descriptors, the $n$-point correlation functions $S_n$ encode the occurrence probabilities of specific $n$-point configurations in the microstructure~\cite{torquato1982microstructure}.
The set of correlation functions up to infinite orders fully characterizes a random field~\cite{torquato1982microstructure, torquato2002random}, and is therefore asymptotically \textit{complete}. While it is empirically shown that some material systems can be represented by \textit{concise} sets of lower-order correlation functions, e.g., metallic alloys, ceramic matrix composites, and certain porous systems~\cite{jiao2013modeling, guo2014accurate, jiao2014modeling, chen2015dynamic, chen2016stochastic, xu2017microstructure, li2018accurate, li2018direct}, there is currently a lack of systematic tools for choosing a concise and nearly complete set of correlations for any particular material system \cite{cheng2021data}. In the case when standard lower-order functions, such as the two-point correlation functions $S_2$, are not sufficient to characterize the system of interest, one can either incorporate non-standard lower-order functions encoding, e.g., clustering or surface information \cite{jiao2009superior}; or employ higher-order functions (e.g., $S_3$) \cite{malmir2018higher}. However, the complexity involved in computing higher-order functions $S_n$ with $n \ge 4$ strongly limits their applications in material modeling.

Recently, we have introduced a set of hierarchical $n$-point polytope functions $P_n$ \cite{chen2019hierarchical, chen2020probing}. The $P_n$ functions provide the probability of finding particular $n$-point configurations associated with regular $n$-polytopes in the material system, and a special sub-set of the standard $n$-point correlation functions $S_n$ that effectively decomposes the structural features in the systems into regular polyhedral basis with different symmetry, and thus, encode partial higher-order correlation information. We have successfully employed time-dependent $P_n$ functions to quantify evolving patterns during thin film deposition \cite{Raghavan2021}, inspired by the work on time-dependent two-point correlation functions \cite{rickman1997impact, rickman2017kinetics, jiao2013modeling, chen2015dynamic}.

Here, we further devise reduced-dimension metrics for effectively measuring the distance between two points (i.e., microstructures) in the microstructure space and quantifying the pathway associated with microstructural evolution, based on the $P_n$ functions. In particular, the $n$-th order metric $\Omega_n$ is defined as the $\mathbb{L}_1$ norm associated with the $P_n$ functions of two distinct microstructures. By choosing a reference initial state (i.e., a microstructure associated with $t_0 = 0$), the $\Omega_n(t)$ set quantifies the evolution of distinct polyhedral symmetries and can in principle capture emerging polyhedral symmetries that are not apparent in the initial state. To demonstrate their utility, we apply the $\Omega_n$ metrics to a 2D binary system undergoing spinodal decomposition to extract the phase separation dynamics via the temporal scaling behavior of the corresponding $\Omega_n(t)$. Moreover, we employ $\Omega_n(t)$ to analyze pattern evolution during vapor-deposition of phase-separating alloy films with different surface contact angles, which exhibit rich evolution dynamics including both unstable and oscillating patterns.

The rest of the paper is organized as follows: In Sec. II, we describe in detail the definition of the polytope functions $P_n$, the associated metric $\Omega_n$, as well as the phase field models for generating microstructural evolution data. In particular, we derive the temporal scaling of $\Omega_n(t)$ and its connection to the temporal scaling of the volume fraction of the evolving system that typically encodes the dynamics signature of the underlying physics. In Sec. III, we present the analysis of a 2D binary system undergoing spinodal decomposition and pattern evolution during vapor-deposition of phase-separating alloy films, using the $\Omega_n(t)$ metrics. In Sec. IV, we provide concluding remarks and discuss potential applications of $\Omega_n(t)$ in establishing quantitative processing-structure-property relationships, as well as in real-time processing control and optimization of complex heterogeneous material systems.

\section{Methods}

\subsection{n-Point polytope functions}



Without loss of generality, consider a heterogeneous material system in $d$-dimensional Euclidean space $\mathbb{R}^d$ with an {\it evolving} binary microstructure in a constant volume $\mathcal{V}$. The snapshot of the microstructure at specific time point $t$ is completely determined by the associated indicator function $\mathcal{L}^{(i)}({\bf x}; t)$, i.e.,
\begin{equation}
\mathcal{L}^{(i)}({\bf x}; t) = \left\{
{\begin{array}{*{20}c}
{1, \quad\quad {\bf x} \in \mathcal{V}_i}\\
{0, \quad\quad {\bf x} \in \overline{\mathcal{V}_i},}
\end{array} }\right.
\end{equation}
where ${\bf x}$ is a position vector in $\mathbb{R}^d$, $i = 1, 2$ is the phase indicator and $\mathcal{V}_i$ indicates regions assocaited with phase $i$. The standard $n$-point correlation function $S_n^{(i)}({\bf x}_1, {\bf x}_2, \ldots, {\bf x}_n; t)$ is defined as \cite{torquato1982microstructure, torquato2002random}
\begin{equation}
S_n^{(i)}({\bf x}_1, {\bf x}_2, \ldots, {\bf x}_n; t) = \langle \mathcal{L}^{(i)}({\bf x}_1; t)\cdot\mathcal{L}^{(i)}({\bf x}_2; t) \ldots \mathcal{L}^{(i)}({\bf x}_n; t)\rangle
\label{eq_Sn}
\end{equation}
where $\langle . \rangle$ denotes ensemble average. $S_n^{(i)}({
\bf X}_n; t)$, where ${\bf X}_n = \{{\bf x}_1, {\bf x}_2, \ldots, {\bf x}_n\}$ provides the probability of finding a specific $n$-point configuration defined by ${\bf X}_n$ in the phase of interest (i.e., phase $i$) at time $t$. In the subsequent discussions, we will drop the phase indicator $i$ for convenience and $S_n({\bf X}_n; t)$ is always associated with the phase of interest.

\begin{figure}
\includegraphics[width=0.325\textwidth,keepaspectratio]{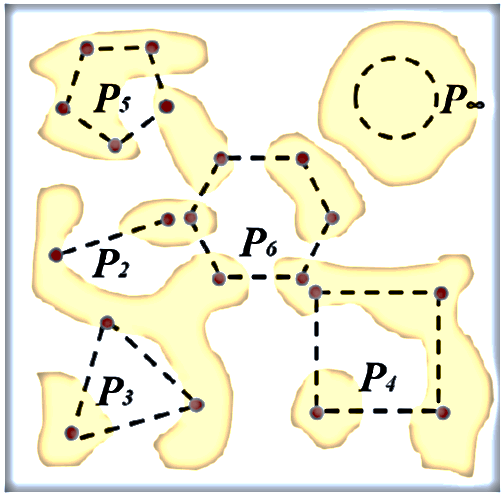}
\caption{Schematic illustration of the $n$-point polytope functions $P_n$.}
\label{fig_1}
\end{figure}

In Refs. \cite{chen2019hierarchical, chen2020probing}, we introduced a special subset of the standard $S_n$, which we referred to as the $n$-point polytope functions $P_n(r; t)$, i.e.,
\begin{equation}
P_n(r; t) = S_n({\bf X}_n~|~{\bf X}_n \in \mathcal{P}(n;r);t)
\label{eq_Pn}
\end{equation}
where $\mathcal{P}(n;r)$ is the set of vertices of a $d$-dimensional regular polytope with $n$ vertices and edge length $r$. $P_n(r; t)$ provides the probability that all of the vertices of a regular $n$-polytope with edge length $r$ fall into the phase of interest when the polytope is randomly placed (both transitionally and rotationally) in the material system at time $t$.
For a statistically homogeneous and isotropic system without long-range orders, $P_n(r=0; t) = \phi(t)$ and $P_n(r\rightarrow\infty; t) = \phi^n(t)$, where $\phi(t)$ is the volume fraction (i.e., probability of a finding a randomly placed point falling into the phase of interest) at time $t$ (see Fig. \ref{fig_1} for illustration). These behaviors allow us to introduce a normalized form of $P_n(r; t)$, i.e.,
\begin{equation}
f_n(r; t) = \frac{P_n(r; t) - \phi^n(t)}{\phi(t)- \phi^n(t)}
\label{eq_fn}
\end{equation}
It is clear from Eq. (\ref{eq_fn}) that $f_n(r=0; t) = 1$ and $f_n(r\rightarrow \infty; t) = 0$.

We note that in $\mathbb{R}^2$, $n$ can take any integer number greater than $d=2$; while in $\mathbb{R}^3$, there are only five regular polyhedra (i.e., the Platonic solids, with $n = 4, 6, 8, 12, 20$) and thirteen semi-regular polyhedra (i.e., the Archimedean solids) that possess uniform lengths for all edges. It has been shown \cite{chen2019hierarchical, chen2020probing, Raghavan2021} that the $P_n$ functions can successively include higher-order $n$-point statistics of the features of interest in the microstructure in a concise, explainable and computationally feasible manner, and can be efficiently computed from given imaging data of the material systems. In addition, the $P_n$ functions effectively ``decompose'' the structural features of interest into a set of ``polytope basis'', allowing one to easily detect any underlying symmetry or emerging features during the structural evolution. Their information content is also investigated via inverse microstructure reconstructions \cite{chen2020probing}.


\subsection{$P_n$-based distance metrics $\Omega_n$  for microstructure space}

The polytope functions $P_n$ allow us to compute corresponding scalar metrics $\Omega_n$ that effectively measure the ``distance'' between to points (i.e., two distinct microstructures) in the microstructure space. Without loss of generality, consider an evolving microstructure ${\cal M}(t)$ that is driven by some external stimuli. We define the distance metric $\Omega_n$ between the microstructure ${\cal M}(t_1)$ and ${\cal M}(t_2)$ as
\begin{equation}
\Omega_n(t) = \frac{1}{N(L)}\sum_{r=0}^L |P_n(r; t_2) - P_n(r; t_1)|,
\label{eq_omega}
\end{equation}
where $|\cdot|$ is the $\mathbb{L}_1$ norm, $t = t_2 - t_1$, $L$ is the edge length of the largest polytope considered and $N(L)$ is the number of different sized polytopes. We note that $\Omega_n$ defined in Eq. (\ref{eq_omega}) quantifies the distinctions between ${\cal M}(t_1)$ and ${\cal M}(t_2)$ with respect to specific $n$-point correlations corresponding to the $n$-polytope configurations. $\mathbb{L}_1$ norm is used here so that the temporal scaling in the volume fraction can be preserved. Similar metrics based on $\mathbb{L}_1$ norm have been recently employed by Lavrukhin et al. to probe the homogeneity conditions implicitly assumed in the study of heterogeneous materials \cite{gerkeprl}.

In the analysis of an evolving microstructure, it is convenient to select a global reference point, e.g., the initial microstructure ${\cal M}(t=0)$. In this case, $\Omega_n(t)$ measures the ``distance'' from the microstructure ${\cal M}(t)$ at time $t$ to the initial microstructure in the microstructure space, i.e.,
\begin{equation}
\Omega_n(t) = \frac{1}{N(L)}\sum_{r=0}^L |P_n(r; t) - P_n(r; t=0)|,
\label{eq_omega2}
\end{equation}
In the subsequent analysis, we will show that although the actual value of the $\Omega_n$ metrics depend on the choice of the global reference point, their temporal scaling behaviors encoding dynamics of the evolving microstructure are independent of the choice of the reference, when the dynamics governing the evolution is temporally homogeneous.

It is also useful to consider the ``distance'' between two successive snapshots of microstructures ${\cal M}(t)$ and ${\cal M}(t-\delta t)$ during the entire evolution process. This allows us to introduce the metric $\delta \Omega_n(t)$, i.e.,
\begin{equation}
\delta \Omega_n(t) = \frac{1}{N(L)}\sum_{r=0}^L |P_n(r; t) - P_n(r; t-\delta t)|,
\label{eq_d_omega}
\end{equation}
We note that $\delta \Omega_n(t)$ is generally different from the differential of $\Omega_n(t)$, i.e.,
\begin{equation}
\delta \Omega_n(t) \neq d\Omega_n(t) = \Omega_n(t) - \Omega_n(t-\delta t),
\end{equation}
unless $P_n(r; t)$ is a monotonic increasing function of $t$ for all $r$.




Although the $\Omega_n$ metrics are introduced based on $P_n$ functions in this work, they can be readily generalized to incorporate the widest class of known spatial correlation functions for the evolving microstructure, e.g., including the lineal-path function $L$ \cite{lu1992lineal}, surface correlation functions $F_{ss}$ and $F_{sv}$ \cite{prager1963interphase}, cluster functions $C_2(r)$ \cite{torquato1988two}, and the pore-size distribution function $P(r)$ \cite{torquato1991diffusion}. Specifically, one only needs to replace the function $P_n$ in Eqs. (\ref{eq_omega2}) and (\ref{eq_d_omega}) with the corresponding correlation function of interest. The meaning of such defined $\Omega$ metric corresponds to the distance between two microstructures as measured with respect to the corresponding correlation function. In addition, although our main focus here is binary microstructures, the $\Omega_n$ metrics can be readily generalized for multi-phase systems. For example, we can separately compute $\Omega^{(i)}_n$ for each phase $i$ and compute the cross-phase $\Omega^{(ij)}_n$ based on the proper cross-correlation functions.




\subsection{Time-dependent $\Omega_n(t)$ and temporal scaling of volume fraction}


The time-dependent $\Omega_n(t)$ metrics defined in Sec. II.B encode information of the evolution dynamics of the material system, which can be assessed from the temporal scaling analysis of these metrics. Here we show how the temporal scaling of the volume fraction for the phase of interest is encoded and can be extracted from the scaling behavior of $\Omega_n(t)$. We note that Eq. (\ref{eq_fn}) allows us to express $P_n$ explicitly in forms of $f_n$ and $\phi$. For simplicity, we first assume that temporal evolution of $P_n(t)$ is solely due to $\phi(t)$ \cite{jiao2013modeling}, i.e.,
\begin{equation}
P_n(t) = [\phi(t) - \phi^n(t)]f_n(r) + \phi^n(t)
\label{eq_Pnf}
\end{equation}
Combine Eq. (\ref{eq_Pnf}) with Eq. (\ref{eq_omega2}), we have
\begin{equation}
\Omega_n(t) = {\cal H}_n\gamma(t) + [1-{\cal H}_n]\gamma_n(t), \label{eq_omega_scaling}
\end{equation}
where
\begin{equation}
{\cal H}_n = \frac{1}{N(L)}\sum_{r=0}^L f_n(r)
\label{eq_H}
\end{equation}
and
\begin{equation}
\gamma(t) = \phi(t) - \phi(0)
\end{equation}
characterizes the temporal scaling of volume fraction and
\begin{equation}
\gamma_n(t) = \phi^n(t) - \phi^n(0)
\end{equation}
It is clear from the above two equations that for higher order $n$, $\gamma_n(t) << \gamma(t)$, and thus the temporal scaling of $\Omega_n(t)$ is mainly dominated by $\gamma(t)$ (i.e., the scaling of volume fraction)
\begin{equation}
\Omega_n(t) \sim {\cal H}_n\gamma(t)
\label{eq_omega3}
\end{equation}
We note that when the scaled function $f_n$ is also time-dependent, we have ${\cal H}_n(t)$ and the scaling behavior of $\Omega_n(t)$ will depends on both ${\cal H}_n(t)$ and $\gamma(t)$. Following the same analysis, we can obtain the temporal scaling behavior for $\delta \Omega_n(t)$, i.e.,
\begin{equation}
\delta \Omega_n(t) = {\cal H}_n\omega(t) + [1-{\cal H}]\omega_n(t), \label{eq_domega_scaling}
\end{equation}
where ${\cal H}_n$ is given by Eq. (\ref{eq_H}) and
\begin{equation}
\omega(t) = \phi(t) - \phi(t-\delta t) = \delta \phi(t)
\end{equation}
and
\begin{equation}
\omega_n(t) = \phi^n(t) - \phi^n(t-\delta t).
\end{equation}
Similar, for higher-order n, the scaling behavior of $\delta \Omega_n(t)$ is mainly determined by $\delta \phi(t)$, i.e.,
\begin{equation}
\delta \Omega_n(t) \sim {\cal H}_n\delta \phi(t).
\label{eq_domega_x}
\end{equation}

\subsection{Phase-field models for microstructrue evolution}


To demonstrate the utility of the $\Omega_n(t)$ for quantifying microstructure evolution, we will employ phase field models to generate 4D data for a variety of evolving material systems. Specifically, we will consider a binary Cahn-Hilliard model with a double-well potential, which has been widely used to simulate phase separate in a simple binary system. First, we will simulate the microstructural evolution of 2D system using the binary model. Next, we will employ a recently developed ternary Cahn-Hilliard model \cite{Raghavan2021} to simulate the temporal evolution of a deposition process of a binary thin film exposed to a vapor phase.

We note that in this subsection, the symbol $\phi$ is used as the order parameter for the phase field models, following the convention in literature, which explicitly depends on the position vectors, i.e., $\phi({\bf x})$. It should not be confused with the phase volume fraction discussed in the previous subsections.

\subsubsection{Binary Cahn-Hilliard Model}

Phase-separations in a 2-phase bulk system are governed by minimizing the total free energy of the system, represented by the equation

\begin{equation}
F = \int_{V}[ f(\phi) + \dfrac{1}{2} \kappa |\nabla \phi|^2 ] dV
\end{equation}
where $\phi$ is a conserved order parameter, representing the composition of the phases, $\kappa$ is the gradient free energy coefficient of the concentration fields and the chemical free energy density is defined by a double-well potential, $f(\phi) = \dfrac{1}{4} W \phi^2 {(1 - \phi)}^2$. $W$ is the well height which penalizes all states other than 0 and 1. The kinetics of phase-separation are simulated by solving the Cahn-Hilliard equation \cite{Cahn1958}, given by

\begin{equation}
\dfrac{\partial \phi}{\partial t} = \nabla.(M \nabla \mu)
\end{equation}
where $\mu$ denotes the chemical potential, and is given by the variational derivative of the free energy functional with respect to the order parameter, $\partial F/\partial \phi$. $M$ denotes the mobility term, which is independent of composition in our study. The temporal evolution of $\phi$ is obtained by incorporating the variational derivative of the free energy functional in the Cahn-Hilliard equation, with the expression,

\begin{equation}
\label{final_eqn}
\dfrac{\partial \phi}{\partial t} = M \nabla^2\left(\dfrac{1}{2}W (2 \phi^3 - 3 \phi^2 + \phi) - \kappa \nabla^2 \phi\right)
\end{equation}
Eq. \ref{final_eqn} is made dimensionless by using reduced variables \cite{Raghavan2020} which are given by: $x^* = x/\Delta x$, $M^* = M/(M_0 k_BT)$, $\nabla^* = (\Delta x)^2 \nabla$, $W^* = W/(k_BT)$, $\kappa^* = \kappa/((\Delta x)^2 k_BT)$, and $t^* = M_0 t/(\Delta x)^2$, where $\Delta x$ is the grid spacing, $M_0$ is an arbitrarily defined temperature-dependent bulk mobility, and $k_B$ is the Boltzmann constant.

\subsubsection{Ternary Cahn-Hilliard Model}

In order to simulate temporal evolution of a binary film exposed to a vapor phase, we adapt a ternary Cahn-Hilliard model for vapor deposition \cite{Cahn1958,Morral1971,CHEN19943503, Raghavan2021}, by assigning field variables to the A-rich and B-rich phases within the film, and the vapor phase within a film-vapor model framework. The evolution is governed by a phenomenological minimization of the free energy functional, given by
\begin{equation}
\begin{array}{c}
F = \int_V N_v \Big[ f(\phi_A, \phi_B, \phi_v) + \kappa_A (\nabla \phi_A)^2 + \\
\kappa_B (\nabla \phi_B)^2 + \kappa_v (\nabla \phi_v)^2 \Big] dV, \; \;
\end{array}
\label{ternary_free_energy_func_eqn}
\end{equation}
where, $N_v$ is the number of molecules per unit volume (assumed independent of composition and position) and $\kappa_i$ ($i = A, B$ and $\nu$) are the gradient energy coefficients. We maintain mass conservation by imposing $\phi_A+\phi_B+\phi_v = 1$. The chemical free energy expression is based on a regular solution, and written as

\begin{equation}
\dfrac{1}{k_BT} f(\phi_{A}, \phi_{B}, \phi_v) = \sum_{i \neq j} \chi_{ij} \phi_i \phi_j + \sum_i \phi_i \mathrm{log}\, \phi_i,
\label{regular_soln_eqn}
\end{equation}
where, $\chi_{ij}$ ($i,j = $A, B, v; $i \neq j$) are the pairwise interaction energies between the components, $k_B$ is the Boltzmann constant, and T, the absolute temperature.

%

The kinetics of phase-separation are obtained via a continuity equation, given by
\begin{equation}
\qquad  \dfrac{\partial \phi_i}{\partial t} = -\nabla\cdot {J_i}'  \quad (i = A, B, v)
\label{temporal_eqn}
\end{equation}
where ${J_i}'$ is the total flux of each component in the system. We adopt a formulation that incorporates the net vacancy flux coupled with a Gibbs-Duhem relation as elaborated by Raghavan et al. \cite{Raghavan2021} and others \cite{kramer1984interdiffusion,huang1995phase,bhattacharyya2003study,sugathan2020phase} to derive the temporal evolution of the A and B-rich phases,

\begin{equation}
\begin{array}{c}
\dfrac{\partial \phi_A}{\partial t} = M_{AA} \nabla^2 \big[(\partial f/\partial \phi_A) - 2\kappa_{AA} \nabla^2 \phi_A - 2\kappa_{AB} \nabla^2 \phi_B\big] \\
+ M_{AB} \nabla^2 \big[(\partial f/\partial \phi_B) - 2\kappa_{BA} \nabla^2 \phi_A -2\kappa_{BB} \nabla^2 \phi_B \big]
\end{array}
\label{ternary_temp_eqn1}
\end{equation}
and
\begin{equation}
\begin{array}{c}
\dfrac{\partial \phi_B}{\partial t} = M_{BB} \nabla^2 [(\partial f/\partial \phi_B) - 2\kappa_{BB} \nabla^2 \phi_B - 2\kappa_{BA} \nabla^2 \phi_B] \\
+ M_{AB} \nabla^2 [(\partial f/\partial \phi_A) - 2\kappa_{AB} \nabla^2 \phi_B -2\kappa_{AA} \nabla^2 \phi_A ],
\end{array}
\label{ternary_temp_eqn2}
\end{equation}
where, $\kappa_{AA} = \kappa_A + \kappa_v$, $\kappa_{BB} = \kappa_B + \kappa_v$, and $\kappa_{AB} = \kappa_{BA} = \kappa_v$ are the gradient parameters. $M_{AA}$ and $M_{BB}$ are the atomic mobilities of A and B atoms in non-A-rich and non-B-rich phases, respectively, while $M_{AB}$ and $M_{BA}$ are mobilities of A atoms in B-rich phase and B atoms in A-rich phase, respectively \cite{huang1995phase,bhattacharyya2003study}. These are related to the diffusion coefficients of the alloying components, $D_i$, via a Nernst-Einstein relation \cite{cogswell2010phase, Raghavan2021},
\begin{equation}
M_{ii} = \dfrac{1}{k_BT}D_i \phi_i(1 - \phi_i) \quad (i,j = A, B, v)
\end{equation}
and
\begin{equation}
M_{ij} = \dfrac{1}{k_BT}D_i \phi_i \phi_j \quad (i,j = A, B, v, \quad i\neq j)
\end{equation}

Eq. \ref{ternary_temp_eqn1} and \ref{ternary_temp_eqn2} are first made dimensionless by using the relation $l^* = (\kappa_i/2k_B T)^{1/2} \Delta x$ and $t^* = (k_B T/M^*_{ii} l^{*2})$ where $l^*$ and $t^*$ are the characteristic length and time, respectively, and $M^*_{ii}$ is the dimensional value of mobility for phases $i = A,B$. The dimensionless form of these equations are then solved via an explicit finite difference scheme for temporal and spatial derivatives.

The evolution of hillocks on the film surface are governed by the interplay of the energies at the interface between A-rich and B-rich phases, and the surface energy of the film. This relationship is encapsulated within Young's equation as

\begin{equation}
\theta = 2 \; \mathrm{cos}^{-1} \left(\dfrac{\sigma_{AB}}{2\sigma_{fv}}\right)
\label{Eq:Young}
\end{equation}

where $\theta$ is the contact angle at the surface, $\sigma_{AB}$ is the interfacial energy between the phase-separated A-rich and B-rich domains and $\sigma_{fv}$ is the energy of the film surface in contact with the vapor phase. $f$ denotes the phase (A-rich or B-rich) at the film surface which is in contact with the vapor phase. The surface energies of both A-rich and B-rich phases in contact with the vapor phase are assumed to be equal. The methodology employed to calculate the contact angles and other model-specific information is available in \cite{Raghavan2021}.

\section{Results}

\subsection{Analysis of binary double-well system in 2D}




To demonstrate the utility of the $\Omega_n$ and $\delta \Omega_n$ metrics, we first apply them to quantify the evolution of a 2D binary system driven by a Cahn-Hilliard model with a double-well potential (see Sec.II.D for details). It is well know that such a system undergoes a rapid phase separation via spinodal decomposition. For this system, it is natural to choose the initial configuration as the reference state to compute $\Omega_n(t)$ via Eq. (\ref{eq_omega2}).



\begin{figure*}[ht]
\includegraphics[width=0.875\textwidth,keepaspectratio]{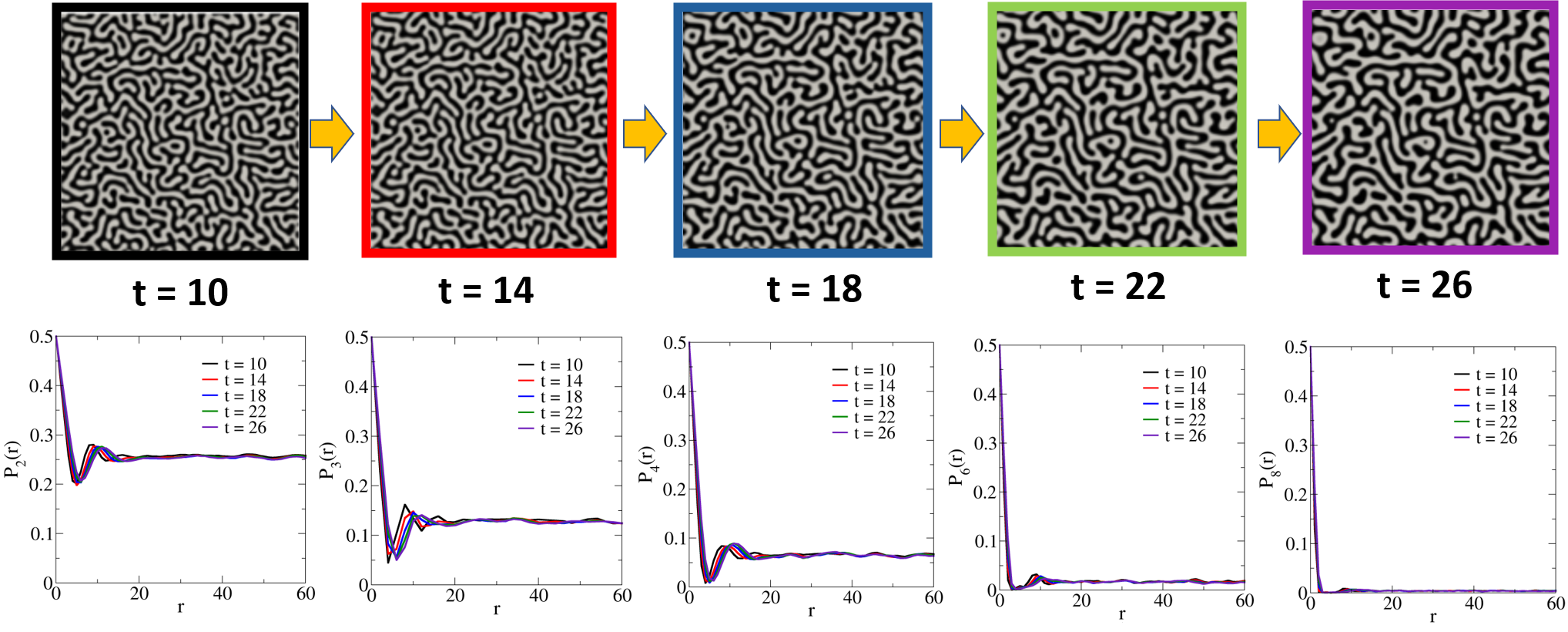}
\caption{Upper panels: Snapshots of the evolving 2D binary system with the double-well potential at selected time points. Lower panels: The associated $n$-point polytope functions $P_n$ at the corresponding time points (indicated by the same color code). The unit of distance $r$ is in pixels.}
\label{fig_2}
\end{figure*}

\begin{figure*}[ht]
\includegraphics[width=0.875\textwidth,keepaspectratio]{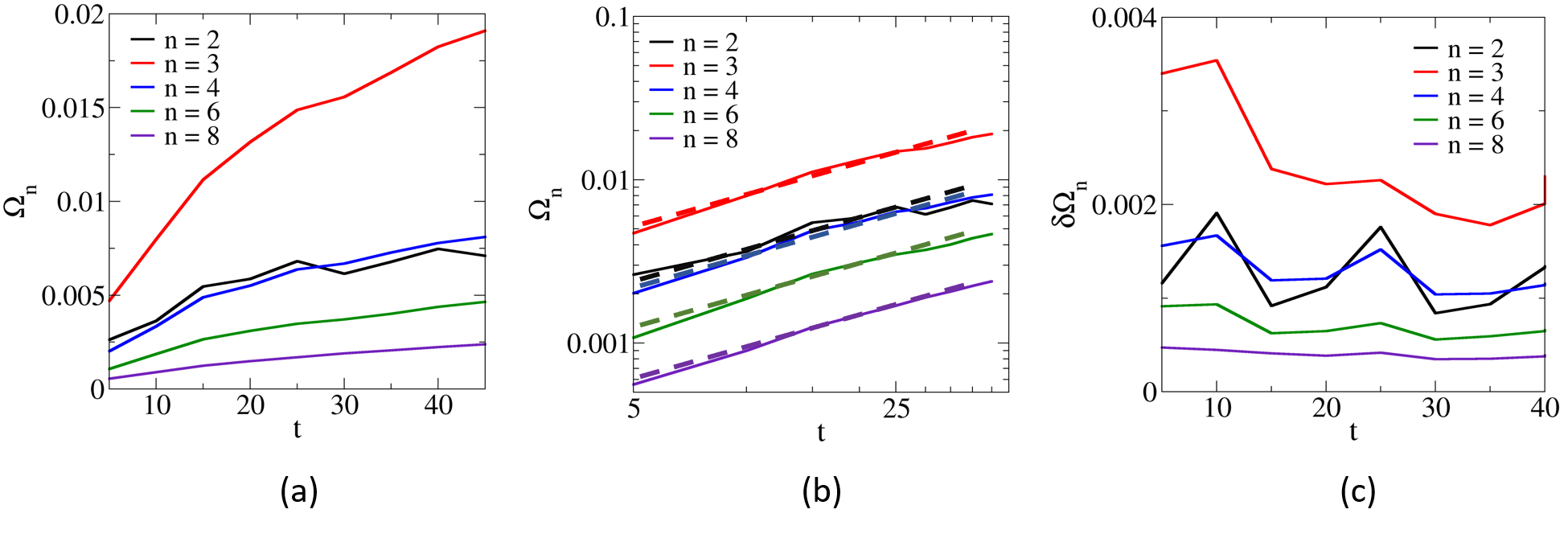}
\caption{$\Omega_n$ metrics for the 2D binary system in both linear scale (a) and log scale (b) with least-square linear fits indicated by dashed lines. $\delta \Omega_n(t)$ for the system is shown in (c).}
\label{fig_3}
\end{figure*}

Fig. \ref{fig_2} upper panels show the selected snapshots of the evolving system, from which one can clearly see the rapid development of the dark phase during the initial stages of the phase separation, which asymptotically slows down. Fig. \ref{fig_2} lower panels shows the $P_n$ (with $n = 2, 3, 4, 6, 8$) for each of these snapshots. A growing length scale can be clearly identified from all of the $P_n$ functions, manifested as slower decay of the functions associated with the snapshots corresponding to the later stages of the phase separation. We note that an estimate of this growing length scale can be obtained by finding the distance corresponding to the first local minimum in $P_2$ \cite{jiao2013modeling}, which clearly shifts to larger distance as the phase separation proceeds. The microstructures of this binary system at all time points are composed of disordered interpenetrating morphology typically seen in a spinodal decomposition, without special symmetry emerging. Therefore, all $P_n$ functions rapidly decay to their corresponding long-range asymptotic values with several oscillations at small $r$. This is distinctly different from the patterns associated with the hillock growth process analyzed in the subsequent section, where patterns with significant 4-fold symmetry emerge during the evolution.

Fig. \ref{fig_3} shows the $\Omega_n$ metrics in both linear scale (a) and log scale (b). It can be seen form Fig. \ref{fig_3}(a) that all $\Omega_n(t)$ rapidly converge to their corresponding long-time asymptotic values. We note that $\Omega_2(t)$ for the system is significantly lower in values compared to $\Omega_3(t)$ and $\Omega_4(t)$. This can be understood from Eq.~(\ref{eq_omega_scaling}): the coefficient ${\cal H}_2$ defined by (\ref{eq_H}) is small since $P_2(r)$ (i.e., $f_2(r)$) is an oscillating function of $r$ with alternative positive and negative values, leading to a small sum $\sum_r f_2(r)$ and thus, small ${\cal H}_2$. Therefore, the scaling of $\Omega_2(t)$ is mainly dominated by its higher order term, i.e., $\Omega_2(t) \sim \gamma_2(t)$, leading to smaller values compared to $\Omega_3(t)$ and $\Omega_4(t)$. Material systems with ${\cal H}_2 = 0$ are called {\it hyperuniform}, which is a recently discovered exotic disordered state in condensed matter systems \cite{xu2017microstructure, jiao2021hyperuniformity}. The log plot in Fig. \ref{fig_3}b shows the similar time scaling behaviors for all $\Omega_n(t)$, confirming the approximation (\ref{eq_omega3}). Based on (\ref{eq_omega3}) we obtain $\Omega_n(t) \sim t^{\alpha}$ where $\alpha \approx 0.645$.



Fig. \ref{fig_3}(c) shows $\delta \Omega_n(t)$ for the system, which reflect the differential change of the system as quantified by different $P_n$ functions as time increases. Based on Eq. (\ref{eq_domega_x}), $\delta \Omega_n(t)$ are mainly determined by the differential change of the volume fraction $\delta \phi(t)$, which approaches 0 towards the later stages of the phase separation (i.e., as volume fraction does not significantly change anymore). This is consistent with the trend for $\Omega_n(t)$ as well, which converge to the long-time asymptotic values towards the later stages of the evolution.






\subsection{Analysis of pattern evolution in thin film deposition}





\begin{figure}[ht]
\includegraphics[width=0.495\textwidth,keepaspectratio]{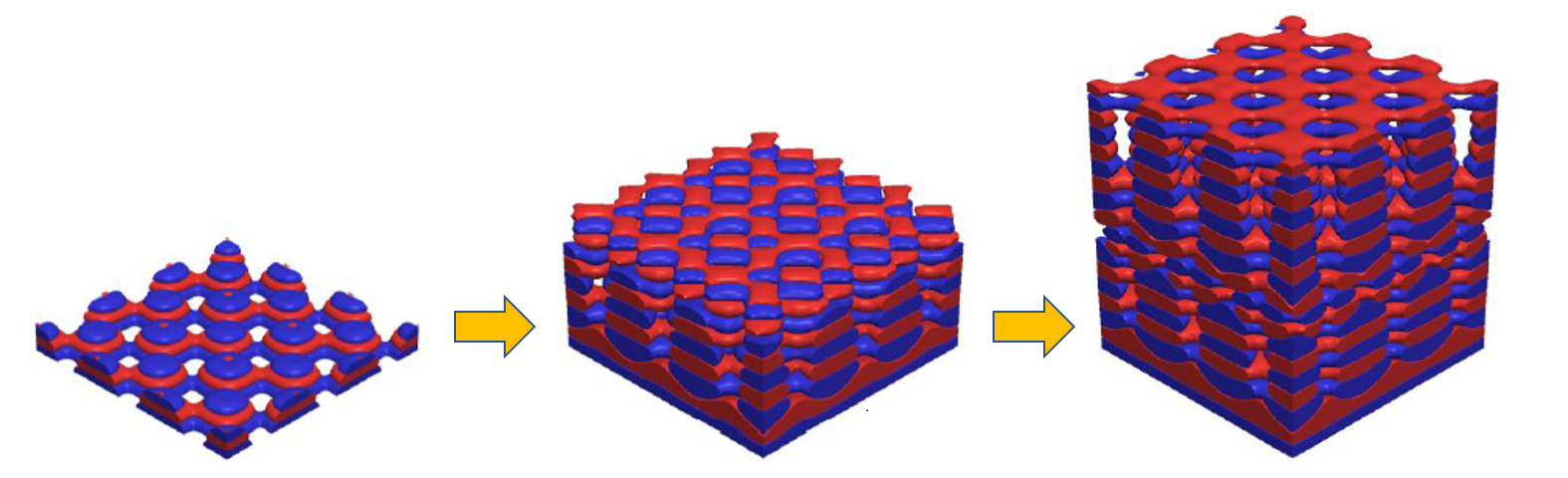}
\caption{Representative snapshots of the growing film containing hillocks with contact angle $\theta = 32^o$ at film-vapor interface (see Ref. \cite{Raghavan2021} for details).}
\label{fig_4}
\end{figure}

With the utilities of $\Omega_n(t)$ and $\delta \Omega_n(t)$ illustrated and verified in the simple binary system discussed in Sec.III.A, we now employ them to analyze the pattern evolution in hillock formation during vapor-deposition of phase-separating alloy films \cite{Raghavan2020, Ankit2019a, raghavan2021multiphysics}. The evolution of the alloy films has been investigated in detail in Ref. \cite{Raghavan2021}, which was simulated via the phase-field model briefly described in Sec.II.D. Figure \ref{fig_4} shows representative snapshots of the growing film containing hillocks with contact angle $\theta = 32^o$ at film-vapor interface (see Ref. \cite{Raghavan2021} for details). Here we will focus on the dynamics of the {\it top slices} of the systems at different times during the film growth. This is motivated by the fact that experimentally the (nano-structured) top surface could be imaged via {\it in situ} characterizations (such as SEM or TEM). We note this is equivalent to take the $x-y$ slices of the thin film configurations associated with different height along $z$ axis starting from the bottom with $z = 0$. Without loss of generality, we only focus on characterizing one of the alloy phases (e.g., the blue phase). We also note that characteristic length scale of the alloy films analyzed here is a few hundred nanometers, and thus, should be referred to as nano-structures. In the following, we will still refer to them as ``microstructure'' with the understanding that such microstructures contain features on nano-scales.

\subsubsection{Pattern evolution with surface contact angle $\theta = 32^o$}

\begin{figure*}[ht]
\includegraphics[width=0.975\textwidth,keepaspectratio]{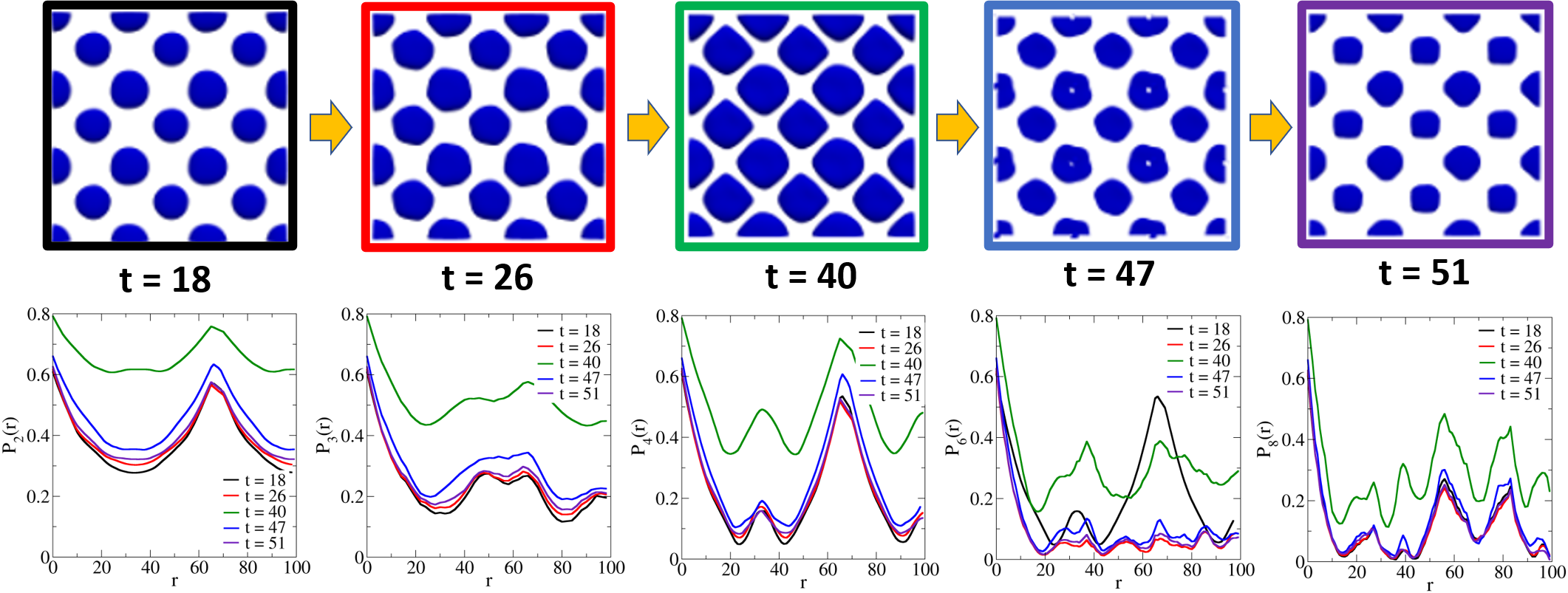}
\caption{Upper panels: Representative snapshots of the top surface patterns associated with the blue phase at different time points during the film growth with contact angle $\theta = 32^o$. Lower panels: The associated $n$-point polytope functions $P_n$ at the corresponding time points (indicated by the same color code). The unit of distance $r$ is in pixels.}
\label{fig_5}
\end{figure*}

\begin{figure*}[ht]
\includegraphics[width=0.875\textwidth,keepaspectratio]{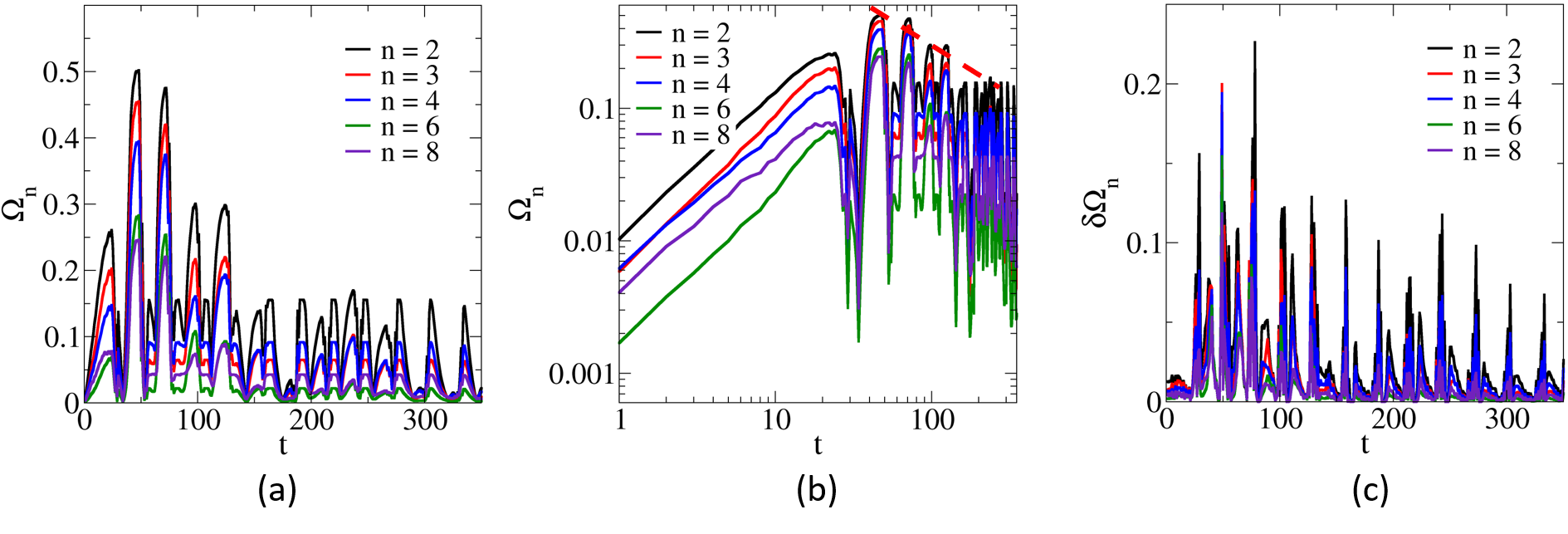}
\caption{$\Omega_n$ metrics for the evolving surface patterns during the film growth with contact angle $\theta = 32^o$ in both linear scale (a) and log scale (b) with least-square linear fit shown as dashed line. $\delta \Omega_n(t)$ for the system is shown in (c).}
\label{fig_6}
\end{figure*}

Figure \ref{fig_5} (upper panels) shows representative snapshots of the top surface patterns associated with the blue phase at different time points during the film growth. It can be clearly seen that the evolving patterns exhibit a four-fold symmetry, which is inherited from the initial configuration \cite{Raghavan2021}. In Ref. \cite{Raghavan2021}, we have shown that after an initial ``transition'' period, the surface patterns for both alloy phases start to oscillate as time proceeds, e.g., one of phases starts to grow from smaller seeds and dominate the pattern which is then gradually taken over by the other phase. Fig. \ref{fig_5} (lower panels) shows the corresponding $P_n$ functions for the snapshots. Different from the simple binary system analyzed in Sec.III.A, the 4-fold symmetry of the patterns is clearly manifested as the strong peaks in both $P_2$ and $P_4$, associated with the same distance $r$. The 4-fold symmetry and the resulting periodicity of the structures also lead to the strong peaks observed in $P_3$, $P_6$ and $P_6$, albeit the distances associated with these peaks are different from those in $P_2$ and $P_4$.


Figure \ref{fig_6} shows the $\Omega_n$ metrics in both linear scale (a) and log scale (b), with the reference pattern chosen at $t = 0$. It can be seen form Fig. \ref{fig_6}(a) that all $\Omega_n(t)$ exhibit almost periodic oscillations for the entire evolution, while the peak values of the oscillations fluctuate during the early stages of the evolution and subsequently converge to a steady value. These features are consistent with the observed dynamics of the top surface pattern of the system: the initial fluctuations of the peaks values correspond to the ``transition'' period of the pattern evolution. On the other hand, the steady peak values correspond to steady oscillations of the top surface patterns resulted from alternating dominant red and blue phases as described above. In addition, the magnitude of $\Omega_n(t)$ at a fixed $t$ decreases as $n$ increases, due to the smaller magnitude of $P_n(r)$ (for $r>0$) as $n$ increases. All $\Omega_n(t)$ exhibit similar time scaling behaviors, as can be seem from the log plot shown in Fig. \ref{fig_6}(b). We note that since $\Omega_n(t)$ are oscillating functions of $t$, we only focus on the temporal scaling of the peak values with respect to the corresponding long-time asymptotic limit $\overline{\Omega}_n(\infty)$, i.e., $|\Omega_n(t) - \overline{\Omega}_n(\infty)|\sim t^{\alpha}$ where $\alpha \approx 0.557 $.




Fig. \ref{fig_6}(c) shows $\delta \Omega_n(t)$ for the system, which exhibits very similar trend as seen in $\Omega_n(t)$. In particular, all $\delta \Omega_n(t)$ exhibit almost periodic oscillations; the peak values of these oscillations fluctuate during the early transition stages of the evolution and subsequently converge to a steady value. A closer inspection reveals the peak positions of $\delta \Omega_n(t)$ correspond to valleys of $\Omega_n(t)$, where the largest rate of change of $\Omega_n(t)$ occurs. These features indicate that the fastest structural dynamics are associated with the valleys of $\Omega_n(t)$, corresponding to patterns reminiscent of the initial configuration at $t = 0$.

\begin{figure*}[ht]
\includegraphics[width=0.75\textwidth,keepaspectratio]{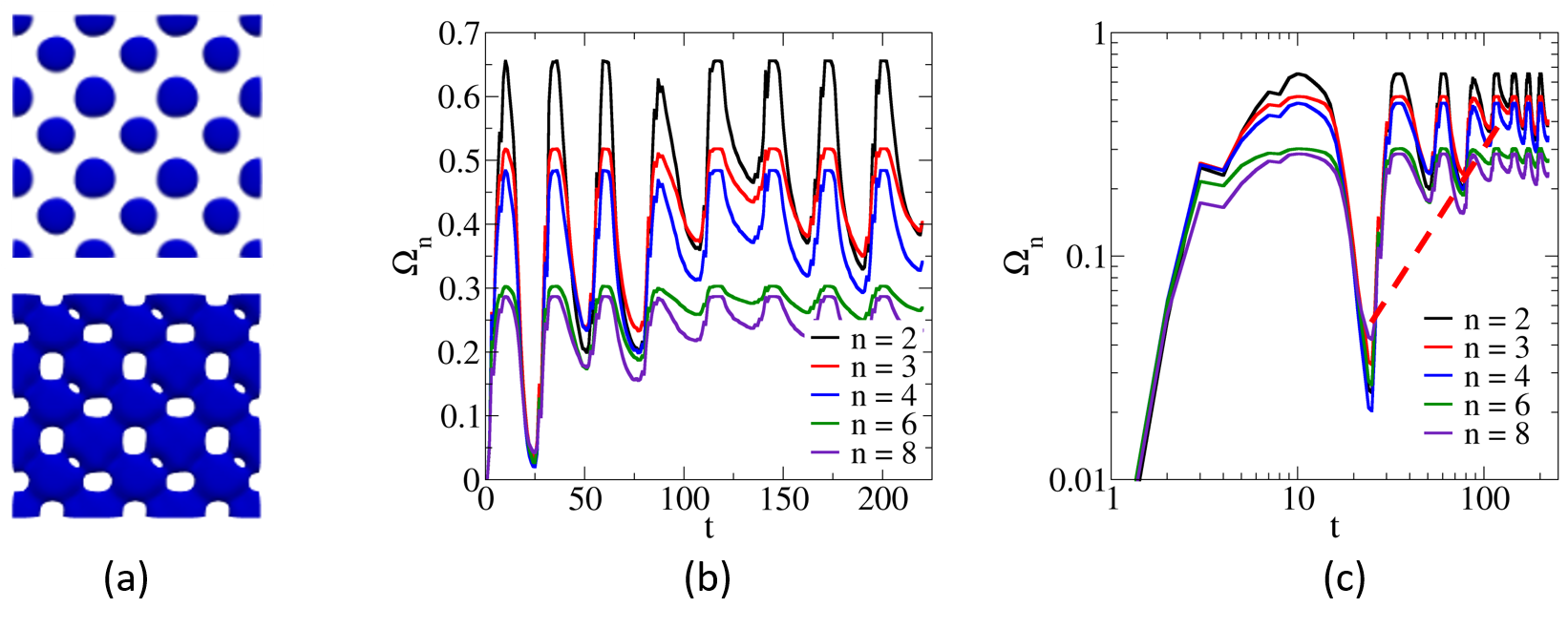}
\caption{The $\Omega^*_n(t)$ metrics for the the evolving surface patterns during the film growth with contact angle $\theta = 32^o$ with a different reference state. (a) respectively shows the original reference (upper panel) and the new reference (lower panel). (b) shows $\Omega^*_n(t)$ in linear scale and (c) shows $\Omega^*_n(t)$ log scale with linear fitting shown as dash line.}
\label{fig_7}
\end{figure*}

For an oscillating system, the choice of initial reference state is not unique. Fig. \ref{fig_7} shows the $\Omega^*_n(t)$ metrics for the system with a different reference state, which is chosen to be the one corresponding to the highest peak of $\Omega_2(t)$ for all $t$, denoted by ${\cal M}^*_0$. Based on the physical interpretation of $\Omega_n$, which is an effective measure of the distance between two microstructures in the material microstructure space, ${\cal M}^*_0$ represents the microstructure that has the largest distance, or in other words, distinct most from, all the other microstructures (patterns) during the evolution. We note that the effective distance between two microstructures can be different based on different $P_n$ measures. These features are all reflected in Fig. \ref{fig_7}(a). In particular, one can clearly see that all $\Omega^*_n(t)$ exhibit coherent converging behavior as $t$ increases, and each $\Omega^*_n(t)$ converges to a set of distinct asymptotic lower and upper bounds for their steady oscillations, indicating different distances to ${\cal M}^*_0$ as measured via different $P_n$. Fig. \ref{fig_7}(b) show the log plot of $\Omega^*_n(t)$, from which we extract the temporal scaling behavior $|\Omega^*_n(t) - \overline{\Omega}^*_n(\infty)| \sim t^{\alpha^*}$ where $\alpha^* \approx 0.594 $, which is consistent with $\alpha \approx 0.557 $ estimated above, and $\overline{\Omega}^*_n(\infty)$ is the long-time asymptotic limit of $\Omega^*_n(t)$. These results indicate that the key behaviors of the system encoded in $\Omega_n$ metrics do not depend on the choice of the reference states.



\subsubsection{Pattern evolution with surface contact angle $\theta = 51^o$}


\begin{figure*}[ht]
\includegraphics[width=0.875\textwidth,keepaspectratio]{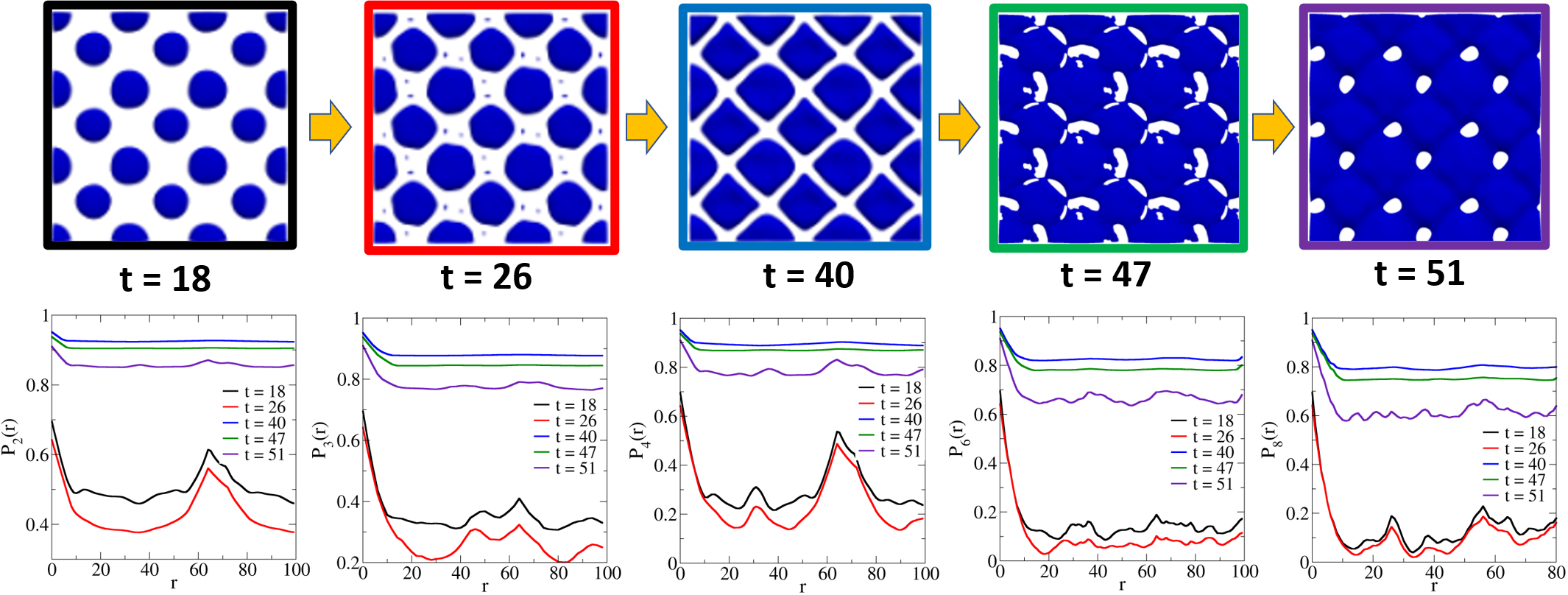}
\caption{Upper panels: Representative snapshots of the top surface patterns associated with the blue phase at different time points during the film growth with contact angle $\theta = 51^o$. Lower panels: The associated $n$-point polytope functions $P_n$ at the corresponding time points (indicated by the same color code). The unit of distance $r$ is in pixels.}
\label{fig_8}
\end{figure*}

\begin{figure*}[ht]
\includegraphics[width=0.875\textwidth,keepaspectratio]{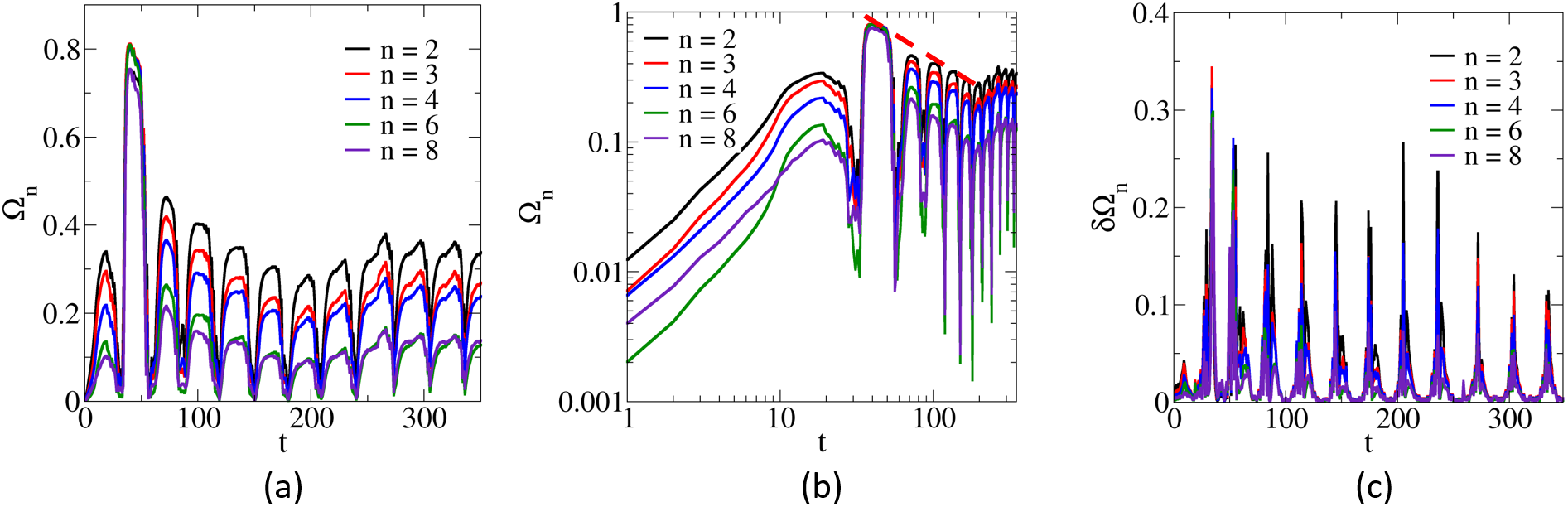}
\caption{$\Omega_n$ metrics for the evolving surface patterns during the film growth with contact angle $\theta = 51^o$ in both linear scale (a) and log scale (b) with least square linear fitting shown as dashed line. $\delta \Omega_n(t)$ for the system is shown in (c).}
\label{fig_9}
\end{figure*}

Following the same procedure, we now employ $\Omega_n(t)$ to characterize the pattern evolution during the thin film growth with contact angle $\theta = 51^o$. Figure \ref{fig_8} (upper panels) shows representative snapshots of the top surface patterns associated with the blue phase at different time points during the film growth; and Fig. \ref{fig_8} (lower panels) shows the corresponding $P_n$ functions at selected time points. Similar to the system with $\theta = 32^o$, the patterns exhibit significant 4-fold symmetry inherited from the initial pattern, which leads to the significant peaks in $P_n$ functions. Fig. \ref{fig_9} shows the $\Omega_n(t)$ in both linear (a) and log scales (b). It can be seen that all $\Omega_n(t)$ exhibit a shorter transition zone and rapidly converge to the steady oscillation stage. We extract the temporal scaling behavior $|\Omega_n(t) - \overline{\Omega}_n(\infty)|\sim t^{\alpha}$ where $\alpha \approx 0.689 $, which is larger than the corresponding scaling parameter for the system with $\theta = 32^o$, indicating faster converging behaviors. The $\delta \Omega_n(t)$ metrics are shown in Fig. \ref{fig_9}(c), which again exhibits very similar trend as seen in $\Omega_n(t)$. The peak positions of $\delta \Omega_n(t)$ also correspond to valleys of $\Omega_n(t)$, indicating the largest rate of change of $\Omega_n(t)$ and the fastest structural dynamics occur at these patterns.


\section{Conclusions and Discussion}


In summary, we have introduced a set of novel reduced-dimension metrics, referred to $\Omega_n$ which are based on the set of hierarchical n-point polytope functions $P_n$, for effectively measuring the distance between two points (i.e., microstructures) in the microstructure space and quantifying the pathway associated with microstructural evolution. By choosing a reference initial state (i.e., a microstructure associated with $t_0 = 0$), the $\Omega_n(t)$ set quantifies the evolution of distinct polyhedral symmetries and can in principle capture emerging polyhedral symmetries that are not apparent in the initial state \cite{chen2019hierarchical}. We have also investigated the temporal scaling behaviors of $\Omega_n(t)$ and showed that the evolution dynamics revealing the physical mechanics of the systems can be extract from the scaling behaviors of $\Omega_n(t)$. To demonstrate their utility, we have applied the $\Omega_n$ metrics to characterize a 2D binary system undergoing spinodal decomposition and extract the evolution dynamics via the temporal scaling behavior of the corresponding $\Omega_n(t)$. We have also employed $\Omega_n(t)$ to quantify pattern evolution during vapor-deposition of phase-separating alloy films with different surface contact angles, which exhibit rich evolution dynamics including both unstable and oscillating patterns.

We note that the $\Omega_n$ metrics are merely a special example of correlation function based distance measures of microstructure space. Similar metrics can be defined for special lower-order functions \cite{jiao2009superior} such as two-point cluster function $C_2(r)$, the surface-surface correlation function $F_{ss}(r)$, the lineal-path function $L(r)$, and the pore-size distribution function $P(r)$, to name but a few. The corresponding $\Omega$ metric quantifies the effective ``distance'' between two microstructures, mainly resulted from the distinction of the structural features quantified by the specific correlation functions. For example, the metric $\Omega_C$ based on the cluster function $C_2$ distinguishes two microstructures based on their degrees of clustering, while these two microstructures may possess identical $P_2$ and $\Omega_2 = 0$ \cite{gommes2012density, gommes2012microstructural}.

The $\Omega_n$ metrics, when combined with {\it in situ} microstructural characterization tools (such as x-ray tomographic microscopy), allow one to quantitatively monitor the structural evolution in real time. During a manufacturing process, this will enable us to apply real-time control of the processing conditions in order to control the microstructure evolution pathway, which is highly desirable for material optimization. In this work, we have used $\mathbb{L}_1$ norm in the definition of $\Omega_n$, which allowed us to extract temporal scaling that characterizes the dynamics of the structural evolution. This can also be generalized to use more sophisticated weighted norms that approximately connect the $P_n$ functions to the physical properties of the material system \cite{chen2019hierarchical}. We will explore these directions in our future work.



\bigskip
\begin{acknowledgments}
This work is supported by National Science Foundation, Division of
Material Research under grant NO. 2020277 (AI Institute: Planning:
Novel Neural Architectures for 4D Materials Science). This
research used computational resources of the Agave Research
Computer Cluster of ASU. Y. J. thanks the generous support from
ASU during his sabbatical leave.
\end{acknowledgments}


\end{document}